\documentclass[11pt,twoside]{article}

\usepackage{asp2004}
\usepackage{epsf}
\usepackage{psfig}
\usepackage{lscape}

\begin{document}
\title{Infrared Properties of AGN in the GOODS Fields: AGN in Stellar
Dominated High-$z$ Galaxies}   
\author{Jeffrey Van Duyne, C. Meg Urry, Eleni Chatzichristou, Ezequiel
Treister, GOODS AGN Team}

\affil{Yale University, P.O. Box 208101, New Haven, CT 06520-8101 USA}

\begin{abstract} 
  We present analysis of spectral energy distributions (SEDs) from
mid-infrared through X-ray of a sample of 420 hard X-ray selected,
z-band and Spitzer/IRAC detected active galactic nuclei (AGN) and AGN
candidates from the GOODS multiwavelength survey.  We fit local
empirical SED templates of both normal and active galaxies to the
rest-frame luminosities calculated from spectroscopic (where
available) and photometric redshifts.  The majority of the optically
stellar-dominated (with early-type galaxy fits) sources are moderately
luminous ($L_{X,2-10 keV} 10^{43}$ erg/s) hard X-ray sources with high X-ray hardness
ratios ($HR > 0.2$), high MIR luminosities and red MIR colors in excess
of a typical stellar dominated elliptical galaxy.  
These sources likely harbor heavily obscured (though Compton-thin) AGN.
The observed ratio of obscured to unobscured AGN has an integrated
mean of $\sim3.4~:~1$ but declines with increasing redshift.  This effect
has been explained by \citet{t04} as an observational bias
triggered by the lack of spectroscopic redshifts at $R > 24$ which are
predominantly higher redshift obscured sources.  

\end{abstract}


  The X-ray background has a far harder X-ray spectrum than the typical unobscured
AGN \citep{main02}.  While a few obscured quasars ($L_{X,2-10
keV}\sim10^{45}$ erg/s) have now been observed
(\citeauthor{norman02}\citeyear{norman02}, XID-S 188),
that population alone does not satisfy the fraction of obscured AGN
predicted by the X-ray background.  Thus, we must also investigate obscured,
moderate luminosity AGN out to $z\sim 1.5-2$, with enough dust
absorption such that the host galaxy dominates the optical and NIR
light with its stellar contribution.  Such AGN would have significant
observed hard X-ray luminosities, and mid-infrared luminosities and
colors in excess of a typical elliptical galaxy indicating an obscured
AGN.

  With the exceptional multiwavelength coverage and depth
afforded by the Great Observatories Origins Deep Survey
(GOODS,\citeauthor{giava04}\citeyear{giava04}), we can uncover a
significant population of the obscured AGN class described above.
Utilizing a sample of 420 $z'$-band matched hard X-ray sources in both
the GOODS-N and South fields \citep{a03}, we fit AGN (obscured and
unobscured) and normal galaxy (spirals and ellipticals) templates to
the broadband rest-frame SEDs from $U$ band
to $Spitzer$ 24\micron/8\micron~(CDF-N and CDF-S fields, respectively)
for sources with spectroscopic and photometric redshifts \citep{barger03,s04,mobash04}.  
\begin{figure}[!ht]
\plotfiddle{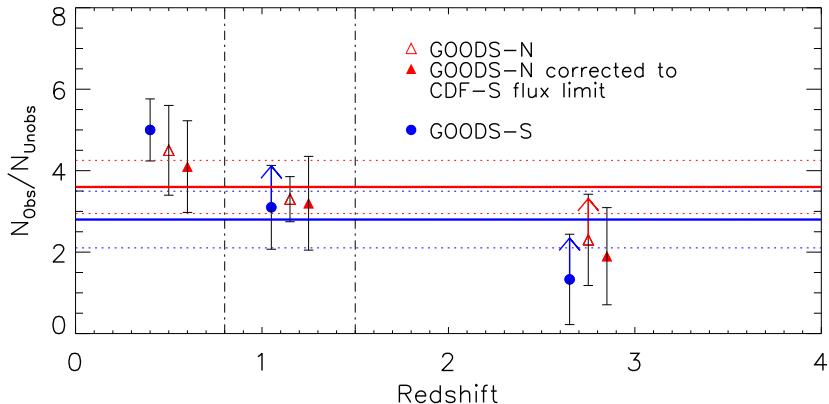}{2in}{0}{62}{62}{-190}{0}
\vspace{-0.3in}
\caption{Observed obscured to unobscured ratio of AGN for GOODS
fields.  Vertical lines represent redshift binning.  Horizontal lines
are global averages at 2.8(3.6) for CDF-N(S) and $3\sigma$ errors.}
\vspace{-0.2in}
\end{figure}

  The template fits result in 82 unobscured AGN, 236 obscured AGN, and
102 other sources (normal ellipticals,,starburst/starforming spirals).  Approximately
half of the unobscured AGN are as weak Seyfert
galaxies in the local universe and have indications of a small amount
of absorption (hardness ratio $> 0.2$ and ($L_{8\micron}/L_{z}$)$ > 0.5$).
Their SEDs fall between a heavily obscured AGN and a Seyfert 1 nucleus,
indicating a partially obscured source ($N_H \sim 10^{21}$).  About 65\% of the obscured AGN are fit
very well to a obscured AGN/Seyfert 2 template from optical to
mid-IR.  The remaining 35\% resemble elliptical galaxies in the
optical, with no obvious point source that would indicate an AGN;
however, hard X-ray luminosities and IR colors reveal the obscured AGN.  Normal
galaxy fits with soft X-ray spectra, $L_X < 10^{42.5}$, and blue IRAC
colors are \emph{bona fide} normal galaxies.  The 24\micron~band in MIPS
shows a slight IR excess (50\% luminosity excess from standard ellipticals) in
25 apparently normal elliptical galaxies in the GOODS-North field.
Also, 24\micron~luminosities identify spiral/starburst
galaxies by [24\micron-3.6\micron] $< -0.5 $ for faint optical sources
that only have photo-$z$.  Based on IRAC/MIPS colors of such sources in the
GOODS-N, we estimate 25\% of the faint
border-line normal galaxies with photo-$z$ in
the GOODS-S will be revealed as definite obscured AGN via 24\micron~
observations.  

  From these fits we compile AGN obscured to unobscured ratios for
three redshift bins (local universe,AGN evolution epoch,AGN formation epoch) for a rough estimate of ratio evolution.  The
observed decline in ratio towards higher redshift is expected if the intrinsic ratio is fixed over all redshifts \citep{t04}, as a result of the
large incompleteness in spectroscopic redshifts for sources with $R >
24$.  Alternate redshifts determinations (X-ray,IR) wil increase the
number of both local and high-$z$, moderate luminosity obscured AGN.  IRS
spectroscopy will verify the nature of the IR bright obscured AGN candidates.


\acknowledgements{This work was supported in part by NASA grant HST-GO-09425.13-A.} 

\vspace{-0.2in}


\end{document}